\documentclass[%
 aip,
 jmp,
 amsmath,amssymb,
 preprint,
 onecolumn, %RC
 amsmath,amssymb,
 showkeys,
 preprint,tightenlines,%
]{revtex4-1}

\usepackage{graphicx}% Include figure files
\usepackage{dcolumn}% Align table columns on decimal point
\usepackage{bm}% bold math
\usepackage[utf8]{inputenc}
\usepackage[T1]{fontenc}
\usepackage{mathptmx}
\usepackage{etoolbox}

\usepackage{color} % statements \color
\usepackage{comment} % http://tex.stackexchange.com/questions/11177/how-to-write-hidden-notes-in-a-latex-file

\usepackage{CJK} % Chinese, Russian \begin{CJK}{UTF8}{song} after begin{document}

\makeatletter
\def\@email#1#2{%
 \endgroup
 \patchcmd{\titleblock@produce}
  {\frontmatter@RRAPformat}
  {\frontmatter@RRAPformat{\produce@RRAP{*#1\href{mailto:#2}{#2}}}\frontmatter@RRAPformat}
  {}{}
}%
\makeatother

\def\today{24 July 2026 with material, accepted}

\def \mod#1{\vert #1 \vert}
\def \ccomma{\raise 2pt\hbox{,}\ } % Le petit livre de TeX page 234
\def \Num{\hbox{\cal N}} 
\def \Den{\hbox{\cal D}} 

\def \Csj{\hbox{Cs}} % cplx affine(sj)
\def \Cdj{\hbox{Cd}} % cplx affine(dj)
\def \Rsj{\hbox{Rs}} % real affine(sj)
\def \Rdj{\hbox{Rd}} % real affine(dj)

\def \Caj{\hbox{Ca}} % cplx affine(aj)
\def \Cbj{\hbox{Cb}} % cplx affine(bj)
\def \Raj{\hbox{Ra}} % real affine(aj)
\def \Rbj{\hbox{Rb}} % real affine(bj)

\def \Nrj{\hbox{Nr}} % num remainder
\def \Drj{\hbox{Dr}} % den remainder

\def \apib{\alpha} % a_j+i b_j
\def \amib{\beta } % a_j-i b_j
\def \Norm{K } 

\begin{document}
%\begin{CJK}{UTF8}{song} % after begin{document}

%\preprint{AIP/123-QED}

\title[NLS rational rogue waves]{Explicit higher order rational rogue waves of the nonlinear Schr\"odinger equation}
% Force line breaks with \\
\author{Robert Conte}
 \altaffiliation[Also at ]{Department of mathematics, The University of Hong Kong,
 Pokfulam, Hong Kong}%Lines break automatically or can be forced with \\
%\author{B. Author}%
 \email{Robert.Conte@cea.fr}
\affiliation{ 
Universit\'e Paris-Saclay, ENS Paris-Saclay, CNRS
\\    Centre Borelli, LRC MESO, F-91190 Gif-sur-Yvette, France
\\ ORCID 0000-0002-1840-5095%\\This line break forced with \textbackslash\textbackslash
}%

% \homepage{http://www.Second.institution.edu/~Charlie.Author.}
%\affiliation{%
%Second institution and/or address%\\This line break forced% with \\
%}%

\date{\today}% It is always \today, today,

\begin{abstract}
The $N$-th order rational rogue wave of the nonlinear Schr\"odinger equation (NLS),
which depends on $2 N-2$ real parameters,
has been shown to be impossible to generate by the nonlinear superposition formula.
We here generate this sequence by a three-term recurrence relation,
each step only requiring the computation of three $N-1$-th order determinants of $2 N-2$ variables,
while the previous method requires two determinants of order $2 N$ in $2 N$ variables.
This allows us to obtain explicitly the seventh wave with its six arbitrary complex parameters.
These very compact expressions open 
the possibility to investigate the possible existence of new patterns
in addition to the already observed ones (concentric rings, polygonal configurations, \dots).
\end{abstract}

% PACS and KEYW HERE

\pacs{
%https://www.aip.org/publishing/pacs/pacs-2010-regular-edition
                        %https://www.aip.org/publishing/pacs/pacs-alphabetical-index		
%02.20.Sv Lie algebras of Lie groups	
%\\ 02.30.Hq Ordinary differential equations
\\ 02.30.Ik Integrable systems 
\\ 02.30.Jr Partial differential equations
% (cosmology, general relativity, nonlinear ODEs, quantum gravity)
\\ 02.30.-f Function theory, analysis						
%\\ 02.40.Dr Euclidean and projective geometries
%\\ 02.40.Hw Classical differential geometry 
}% PACS, the Physics and Astronomy Classification Scheme.

%Use showkeys class option if keyword
															
\maketitle

\tableofcontents
%\vfill\eject

% ============================================================================
\section{Introduction} 

The nonlinear Schr\"odinger equation (NLS)
\begin{eqnarray}
& &
i u_T +p u_{XX} + q \mod{u}^2 u=0,
\label{eqNLS}
\end{eqnarray} 
with $p$ and $q$ real constants,
admits a remarkable solution found by Peregrine \cite{Peregrine-1983},
\begin{eqnarray}
& &
%u(X,T)=\sqrt{-\omega/q}\left(1-4 \frac{1 -2 i \omega T}{1+4 \omega^2 T^2 - 2 (\omega/p) X^2 }\right) e^{-i \omega T},
 u(X,T)=\sqrt{ \omega/q}\left(1-4 \frac{1 +2 i \omega T}{1+4 \omega^2 T^2 + 2 (\omega/p) X^2 }\right) e^{ i \omega T},
\omega \in \mathbb{R},
\label{eqPeregrine-full}
\end{eqnarray} 
whose main property, at least in the focusing r\'egime $p q >0$,
is a high maximal amplitude above its background,
\begin{eqnarray} & &
\frac{|u(0,0)|}{|u(\infty,T)|}=3,
\end{eqnarray} 
making it a prototype of ``freak wave''.

\textit{Remark}.
Under the transformation $u(X,T)=v(X,T) e^{ i \omega T}$,
the solution (\ref{eqPeregrine-full}) becomes a rational solution of the equivalent equation
\begin{eqnarray}
& &
i v_T +p v_{XX} + (q \mod{v}^2 - \omega) v=0,
\label{eqNLS-equivalent}
\end{eqnarray} 
therefore we will simply call the solution (\ref{eqPeregrine-full}) ``rational'' instead of ``quasi-rational''.
\medskip

The invariance of (\ref{eqNLS}) under a Galilean transformation
\begin{eqnarray} & &
\forall c \in \mathbb{R}:\ (u,X,T) \to (u e^{i(2 c X- c^2 T)/(4 p)},X-cT,T)
\end{eqnarray}
and under the scaling 
$\forall k \in \mathbb{R}:\ (u,X,T) \to (k u,  X/k,T/k^2)$
%and under a phase shift 
%$\forall \varphi \in \mathbb{R}:\ u \to u e^{i \varphi}$
allows one to adopt the convenient normalization 
$p=1, q=1/2, \omega= 1/2$, 
i.e.~to represent (\ref{eqPeregrine-full}) as
\begin{eqnarray}
& &
u(X,T)=\left(1-4 \frac{1 + i  T}{1+X^2 + T^2 }\right) e^{i T/2}.
\label{eqPeregrine}
\end{eqnarray}

Starting from the plane wave $u_0$,
the sequence of rational rogue waves is defined as
\begin{eqnarray}
& & {\hskip -15.0truemm}
\left\lbrace
\begin{array}{ll}
\displaystyle{
u_0=e^{i T/2},
}\\ \displaystyle{
u_1=\left(1-4 \frac{1 + i  T}{1+X^2 + T^2 }\right) e^{i T/2},
}\\ \displaystyle{
u_N=\frac{\Num_N(X,T)}{\Den_N(X,T)} e^{i T/2},
}
\end{array}
\right.
\label{eq-N-0-1} 
\end{eqnarray}
with $\Num_N$ (complex) and $\Den_N$ (real) relatively prime polynomials of $X,T$,
	having the same global degree $N(N+1)$ in $(X,T)$
	and depending on $2 N-2$ arbitrary real parameters \cite{Dubard-These-2010,DGKM-2010},
	here denoted $a_j,b_j$.
These two polynomials are normalized by
their value when $X$, $T$ and all the $2 N-2$ real parameters vanish,  
\begin{eqnarray}
& & \Num_N(0,0)=(-1)^N (2 N+1), \Den_N(0,0)=1. % 1,-3,5,-7,9,...
\label{eq-unique}
\end{eqnarray}

In order to build this $N$-th order rational rogue wave,
several methods have been used, which we now summarize.

\begin{enumerate}

	\item After expanding a periodic generating function \cite{EKK1986}
	in powers of the frequency
	(see \cite[\S 2]{DGKM-2010} for full details),
	the limit when the frequency goes to the $\omega$ in (\ref{eqNLS-equivalent})
	provides a representation of the $N$-th rational rogue wave
	as the quotient of two Wronskians of order $2 N$.
	
	\item
	By applying the Darboux transformation to the plane wave $u_0$ \cite{AEK1987},
one obtains a trigonometric solution 
whose rational limit is the Peregrine wave $u_1$.
Iterating the process yields the $N$-th  rational rogue wave as 
the ratio of two determinants of order $2 N$.
% The rational limit should be taken with care,
% otherwise one might set to zero the $2 N-2$ arbitrary real parameters.
	
	\item The finite-gap quasi-periodic solution is first computed \cite{Its-Rybin-Salle-1988.NLS-rogue}
	as the ratio of two determinants of order $2 N$,
	then one takes its rational limit.
	
	\item 
	The order of the two determinants can be divided by two
if one uses the Hirota formalism \cite{Ohta-Yang-2012},
but the two resulting $N$-th order determinants require additional computations
because they are of the type of Gram.
\end{enumerate}

The cost of all these methods (computation of two determinants of order $2 N$ or $N$
depending on the $2 N$ variables $X,T,a_j,b_j$)
grows enormously with $N$.

% ============================================================== NLSF

In principle, there exists another method to obtain the $N$-th rational wave,
this is the nonlinear superposition formula (NLSF) \cite{Lamb1974,RogersShadwick} of NLS,
an algebraic relation between four solutions of NLS,
which is essentially derived from the Darboux and B\"acklund transformations
by an elimination process.
However, as observed in \cite{AAS2009},
the Darboux transformation fails to generate the sequence of $N$-th order rational rogue waves,
therefore the NLSF cannot be used for this purpose.

Consequently,
the explicit expressions of $u_N(X,T)$ depending on $2 N-2$ arbitrary real parameters
could only be obtained for 
$N=2$ by Dubard and Matveev \cite{Dubard-These-2010} \cite{Dubard-Matveev-N-2-2011}, 
% (later in \cite{Gaillard-NLS-N-2-2011}) (2 arbitrary constants).
and for
$N=3$ and $N=4$ by the same authors \cite{Dubard-Matveev-2013-NLS-N2-N3-N4}. 
%(later in \cite{Gaillard-NLS-N-3-2013}) (4 arbitrary constants).
Using computer algebra,
Gaillard then obtained particular cases of
$N=5$ (with 3 constraints on the 8 parameters) \cite{Gaillard-NLS-N-5partial-2013},  
% preprint (14050 pages, how many terms in denom?) hal-00819359
the general $N=6$ solution 
(however ``too monstrous to be published'') \cite{Gaillard-NLS-N-6partial-2014}, 
then, when all $2N-2$ parameters are zero, the orders up to $N=10$ included \cite{Gaillard-NLS-N-10partial-2015},
and finally 12 one-parameter particular solutions for $N=13$
\cite{Gaillard-NLS-N-13partial-2017}.
This was sufficient to exhibit quite a number of regular patterns of rogue waves,
depending on the values of the parameters.

This lack of an explicit expression retaining all $2N-2$ parameters in the $N$-th order rogue wave,
$N \ge 5$,
inhibits the possible discovery of new patterns.

In the present paper,
we build a representation of the $N$-th order rational wave
by much shorter elementary polynomials,
thus making easier the analytic investigation of possibly new patterns.
This is essentially achieved by considering the dependence of $u_N$ not on $X,T$
but on the two complex conjugate parameters $a_{N-1} \pm i b_{N-1}$.
This allows us to write explicitly, in a supplementary data file,
all these elementary polynomials up to $N=7$ included. 

The paper is organized as follows.
In section \ref{section-Matrices},
we first define a canonical representation of the two Wronskians allowing one
to reduce the order $2 N$ of the determinants by a factor two.
In section \ref{section-Recurrence},
after defining two affine functions of $a_{N-1} + i b_{N-1}$
and two other                       of $a_{N-1} - i b_{N-1}$,
we build two identical three-term recurrence relations,
one for the sequence $\Num_N(X,T)$,
the other for the sequence $\Den_N(X,T)$,
thus drastically reducing the cost of the computation.
In section \ref{section-other-recurrences},
we recall other recurrence relations for NLS
and explain why they fail in our case.
Finally, 
section \ref{section-Other} defines an even shorter representation
of the $N$-th rational rogue wave,
which could be linked to some yet undiscovered NLSF.

In the Appendix \ref{Appendix-First-few}, we list all the above mentioned short polynomials
up to order $N=7$ included with their $N-1$ arbitrary complex parameters.

% ============================================================================
\section{Canonical representation of the Wronskians} 
\label{section-Matrices}

We won't repeat here the construction of the two Wronskians,
since full details can be found in \cite{Gaillard-NLS-N-2-2011}.
Let us denote $M_1$ and $M_3$ 
the matrices of order $2 N$ whose Wronskians, defined in \cite[Eqs.~(17), (22)]{Gaillard-NLS-N-2-2011},
characterize the $N$-th rational wave,
\begin{eqnarray}
& & u_N=\frac{\Num_N(X,T)}{\Den_N(X,T)} e^{i T/2}, \Num_N(X,T)=\det M_3, \Den_N(X,T)=\det M_1.
\label{eq-uN-M3-M1}
\end{eqnarray}

To take an example,
their expressions for $N=2$ are
\begin{eqnarray}
& & {\hskip -8.0truemm}
%M_1=
%[[[X+I T,-2 X-6 I T-2 a1+2 I b1+1/3 (-X-I T)^3,0    ,2],
%  [1    ,-(-X-I T)^2                          ,0    ,-2 X+2 I T],
%  [0    ,-2 X-2 I T                           ,-1   ,(X-I T)^2],
%  [0    ,-2                                   ,X-I T,-2 X+6 I T-2 a1-2 I b1-1/3 (X-I T)^3]]] 
M_1=\begin{pmatrix} % den
   X+i T & -2 \amib_1+p_{1,12} & 0    & 2 \cr 
  1      &  -(X+i T)^2                          & 0    &-2 (X-i T) \cr
  0      &-2 (X+i T)                            &-1    &   (X-i T)^2\cr
  0      &-2                                    &X-i T & -2 \apib_1+\overline{p_{1,12}}\cr 
\end{pmatrix},
\nonumber\\ & & {\hskip -8.0truemm}
%M_3=[[X+I T+2,-2 X-6 I T-4/3-2 a1+2 I b1+1/3 (-X-I T-2)^3,0,2],
%[1,-(-X-I T-2)^2,0,-2 X+2 I T+4],
%[0,-2 X-2 I T-4,-1,(X-I T-2)^2],
%[0,-2,X-I T-2,-2 X+6 I T+4/3-2 a1-2 I b1-1/3 (X-I T-2)^3]]
M_3=\begin{pmatrix} % num
  X+i T+2 & -2 \amib_1+p_{3,12}-4/3 & 0 & 2 \cr 
  1       & -(X+i T+2)^2 & 0  & -2 (X- i T-2) \cr
  0       & -2 (X+i T+2)  & -1 & (X-i T-2)^2\cr
  0       & -2 & X-i T-2  & -2 \apib_1+\overline{p_{3,12}}+4/3 \cr 
	\end{pmatrix},
\label{eq-N-2-matrices-before}	
         \\ & & {\hskip -8.0truemm} p_{1,12}=-2 X-6 i T-1/3 (X+i T)^3,
\nonumber\\ & & {\hskip -8.0truemm} p_{3,12}=-2 X-6 i T-1/3 (X+i T-2)^3,
\nonumber
\end{eqnarray}
Because of their common origin as Wronskians,
they have the same analytic structure,
and their elements only differ by additive constants in their factors.
Moreover,
the large number $2 N^2-2 N$ of null elements in each matrix indicates a possible
reduction of the order of the determinants.

This is indeed the case.
After $2 N$ transpositions of rows, these matrices become
\begin{eqnarray}
& & {\hskip -8.0truemm}
%Blocksden:=
%[1,-(X+I T)^2,0,-2 X+2 I T],
%[0,-2,X-I T,-2 X+6 I T-2 a1-2 I b1-1/3 (X-I T)^3],
%[0,-2 X-2 I T,-1,(X-I T)^2],
%[X+I T,-2 X-6 I T-2 a1+2 I b1-1/3 (X+I T)^3,0,2]
\widetilde M_1=\begin{pmatrix} 
  1 & -(X+i T)^2 & 0 & -2 (X- i T) \cr
  0 & -2 & X-i T & -2 \apib_1+p_{1,44} \cr
  0 & -2 (X+ i T) & -1 & (X-i T)^2 \cr
  X+i T & -2 \amib_1+p_{1,12} & 0 & 2 \cr 
\end{pmatrix},
\nonumber\\ & & {\hskip -8.0truemm}
%Blocksnum:=
%[1,-(X+I T-2)^2,0,-2 X+2 I T+4],
%[0,-2,X-I T-2,-2 X+6 I T+4/3-2 a1-2 I b1-1/3 (X-I T-2)^3],
%[0,-2 X-2 I T-4,-1,(X-I T-2)^2],
%[X+I T+2,-2 X-6 I T-4/3-2 a1+2 I b1-1/3 (X+I T-2)^3,0,2]
\widetilde M_3=\begin{pmatrix} 
   1 & -(X+i T-2)^2 & 0 & -2 (X- i T-2) \cr 
   0 & -2 & X-i T-2 & -2 \apib_1+p_{3,44} \cr
   0 & -2 (X+i T+2) & -1 & (X-i T-2)^2 \cr
   X+i T+2 & -2 \amib_1+p_{3,12} & 0 & 2 \cr 
	\end{pmatrix},
\label{eq-N-2-matrices-Blocks}
\end{eqnarray}
i.e.~four blocks of $N$-th order matrices
\begin{eqnarray}
& & \begin{pmatrix}A & B \cr C & D \cr\end{pmatrix},
\label{eq-blocks}
\end{eqnarray}
in which $A$ and $D$ are upper triangular and nonsingular.
The classical formula
\begin{eqnarray}
%det(A B C D)=det(A)*det(D-C*A^{-1}*B)
& & \det \begin{pmatrix}A & B \cr C & D \cr\end{pmatrix}= (\det A) \det(D-C A^{-1} B),
\label{eq-det-blocks}
\end{eqnarray}
in which the off-diagonal block matrices $B$ and $C$ may be rectangular,
then reduces the $2 N$-th order determinants to $N$-th order ones.

Such a reduction of determinants order by a factor two was also achieved
by Ohta and Yang \cite{Ohta-Yang-2012},
but their determinants, of the type of Gram, have a different structure;
in particular, their $M_3$ is Hermitian, which is not the case here.

% ============================================================================
\section{Recurrence relation for the $N$-th rational wave} 
\label{section-Recurrence}

In the first wave Eq.~(\ref{eq-N-0-1}),
the denominator $\Den_1$ is the sum of three squares,
also equal to one square plus a product of two factors.
Because of the similar analytic structure of the two Wronskians,
this is also the case of the numerator,
\begin{eqnarray}
%u1RWter:=(1+X**2+(T-2*I)**2)/(1+X^2+T^2) *exp(I*T/2);
& & 
\Num_1=1+X^2+(T-2 i)^2=(X+i T+2)(X-i T-2)+1, \Num_1(0,0)=-3, 
\nonumber \\ & & 
\Den_1=1+X^2+T^2=(X+i T) (X-i T) +1, \Den_1(0,0)=1,
\end{eqnarray}
the three squares only differing from those of the denominator by constants.
This property holds true for any $N$, as we now show.

In order to achieve this goal, one should not look at the dependence of $\Num_N, \Den_N$ on $(X,T)$,
but on the two real parameters $a_{N-1}$ and $b_{N-1}$, 
or, better, on the two equivalent complex conjugate parameters $\apib_{N-1}, \amib_{N-1}$,
\begin{eqnarray}
& & a_j+i b_j = \apib_j, a_j-i b_j = \amib_j.
\label{eq-def-param}
\end{eqnarray}

As seen on (\ref{eq-N-2-matrices-Blocks}),
for $N \ge 2$, both polynomials $\Num_N=\det M_3$ and $\Den_N=\det M_1$ are bilinear functions of 
the complex conjugate parameters $(\apib_{N-1}, \amib_{N-1})$,
canonically representable by eight polynomials,
\begin{eqnarray}
& &
\left\lbrace
\begin{array}{ll}
    \displaystyle{\Num_N= \frac{\Csj_{N-1} \Cdj_{N-1} + \Nrj_{N}}{\lambda_{N-1}},
}\\ \displaystyle{\Den_N= \frac{\Rsj_{N-1} \Rdj_{N-1} + \Drj_{N}}{    \mu_{N-1}}, 
}
\end{array}
\right.
\end{eqnarray}
in which the two fractions reduce to polynomials.
Four of these eight polynomials can be chosen as the partial derivatives of the two matrices
(note that two of them are complex conjugate).
Introducing the pure number $K_N^2$ as the product $(\det A)(\det B)$ of the eigenvalues of
the two $N$-th order diagonal blocks of Eq.~(\ref{eq-blocks}),
\begin{eqnarray}
& &
\Norm_N=2! 4! \dots (2N-2)!,
\end{eqnarray}
these four polynomials are defined as
\begin{eqnarray}
& &
\left\lbrace
\begin{array}{ll}
    \displaystyle{\Rsj_{N-1}=\Norm_N^{-2} \det \partial_{\amib_{N-1}} M_1=\hbox{ affine function of } \apib_{N-1},
}\\ \displaystyle{\Rdj_{N-1}=\Norm_N^{-2} \det \partial_{\apib_{N-1}} M_1=\hbox{ affine function of } \amib_{N-1} = \overline{\Rsj_{N-1}},
}\\ \displaystyle{\Csj_{N-1}=\Norm_N^{-2} \det \partial_{\amib_{N-1}} M_3=\hbox{ affine function of } \apib_{N-1},
}\\ \displaystyle{\Cdj_{N-1}=\Norm_N^{-2} \det \partial_{\apib_{N-1}} M_3=\hbox{ affine function of } \amib_{N-1},
}
\end{array}
\right.
\label{eq-def-squares} 
\end{eqnarray}
leaving only four polynomials to be determined.

When one eliminates the two parameters $\apib_{N-1}$, $\amib_{N-1}$
between $\det M_3$ (resp.~$\det M_1$) and the product $\Csj_{N-1} \Cdj_{N-1}$ (resp.~$\Rsj_{N-1} \Rdj_{N-1}$),
it turns out that the result $\Nrj_{N}$ (resp.~$\Drj_{N}$) is the square of $\Num_{N-1}$ (resp.~$\Den_{N-1}$), more precisely,
\begin{eqnarray}
& & (N\ge 2)
\left\lbrace
\begin{array}{ll}
%Num3:=factor(subs(rtoab,(Csj*Cdj+Num2^2)/Num1)):
%Den3:=factor(subs(rtoab,(Rsj*Rdj+Den2^2)/Den1)):
    \displaystyle{\Den_N = \det M_1 = \frac{\Rsj_{N-1} \Rdj_{N-1}+ \Den_{N-1}^2}{\Den_{N-2}}\ccomma
}\\ \displaystyle{\Num_N = \det M_3 = \frac{\Csj_{N-1} \Cdj_{N-1}+ \Num_{N-1}^2}{\Num_{N-2}}\cdot
}
\end{array}
\right.
\label{eq-recurrence-N} 
\end{eqnarray}

This system defines two independent three-term recurrence relations,
one for the numerators   $\Num_N$,
one for the denominators $\Den_N$.

The first terms are
\begin{eqnarray}
& & (N=2)
\left\lbrace
\begin{array}{ll}
    \displaystyle{\Den_2 = \det M_1 = \frac{\Rsj_1 \Rdj_1 + \Den_1^2}{\Den_0}\ccomma 
    \Den_0=1, \Norm_2=2! = 2, 
}\\ \displaystyle{\Num_2 = \det M_3 = \frac{\Csj_1 \Cdj_1 + \Num_1^2}{\Num_0}\ccomma \Num_0=1,
}
\end{array}
\right.
\end{eqnarray}
\begin{eqnarray}
& & (N=3)
\left\lbrace
\begin{array}{ll}
    \displaystyle{\Den_3 = \det M_1 = \frac{\Rsj_2 \Rdj_2 + \Den_2^2}{\Den_1}\ccomma \Norm_3=2! 4! = 48,
}\\ \displaystyle{\Num_3 = \det M_3 = \frac{\Csj_2 \Cdj_2 + \Num_2^2}{\Num_1}\ccomma
}
\end{array}
\right.
\end{eqnarray}
\begin{eqnarray}
& & (N=4)
\left\lbrace
\begin{array}{ll}
%Den4:=simplify(subs(rtoab,(Rsj*Rdj+Den3^2)/Den2)):
%Num4:=simplify(subs(rtoab,(Csj*Cdj+Num3^2)/Num2)):
    \displaystyle{\Den_4 = \det M_1 = \frac{\Rsj_3 \Rdj_3 + \Den_3^2}{\Den_2}\ccomma \Norm_4=2! 4! 6!= 34560,
}\\ \displaystyle{\Num_4 = \det M_3 = \frac{\Csj_3 \Cdj_3 + \Num_3^2}{\Num_2}\cdot
}
\end{array}
\right.
\end{eqnarray}

There only remains to compute the 
four polynomials 
	$\Csj_{N-1}$, $\Cdj_{N-1}$, $\Rsj_{N-1}$, $\Rdj_{N-1}$
	as the determinants of the respective partial derivatives 
	$\partial_{\amib_{N-1}} {M_3}$,
	$\partial_{\apib_{N-1}} {M_3}$,
	$\partial_{\amib_{N-1}} {M_1}$,
	$\partial_{\apib_{N-1}} {M_1}$.
	Since the last two polynomials are complex conjugate, only three of them need be computed.
	This is an easy and cheap task because
	each variable $\apib_{N-1}$, $\amib_{N-1}$ appears only once in the matrices,
	therefore these four partial derivatives are not a sum of determinants but a single determinant;
	moreover, the derivation operation creates $2 N-1$ zeroes which, added to the existing zeroes of (\ref{eq-N-2-matrices-Blocks}),
	recreate an upper left block, this time of lower order $N-1$, which is upper triangular and nonsingular.
	Let us again take the example $N=2$ and compute $\Rsj_{N-1}$, proportional to $\partial_{\amib_{N-1}} {M_1}$.

One starts from	
\begin{eqnarray}
& & {\hskip -8.0truemm}
%[[    1,-(-X-I T)^2                   ,    0,-2 X+2 I T],
% [    0,-2                            ,X-I T,-2 X+6 I T-2 s1-1/3 (X-I T)^3],
% [    0,-2 X-2 I T                    ,   -1,(X-I T)^2],
% [X+I T,-2 X-6 I T-2 d1+1/3 (-X-I T)^3,    0,2]]
\det \partial_{\amib_{1}} \widetilde M_1=\det \begin{pmatrix} 
  1     & -(X+i T)^2          & 0     & -2 (X- i T) \cr
  0     & -2                  & X-i T & -2 \apib_1 - 2 X + 6 i T -(X-i T)^3/3 \cr
  0     & -2 (X+ i T)         & -1    & (X+i T)^2 \cr
  0     & -2                  & 0     & 0\cr 
\end{pmatrix}.
\end{eqnarray}
Then, the $2N-1$ zeros of column 1 and line $2 N$ make this determinant equal
to the product of the two nonzero elements by their minor of order $2 N-2$,
\begin{eqnarray}
& & {\hskip -8.0truemm}
\det \partial_{\amib_{1}} \widetilde M_1=-2 \det \begin{pmatrix} 
%[[X-I T,-2 X+6 I T-2 s1-1/3 (X-I T)^3],[-1,(X-I T)^2]]
 X-i T & -2 \apib_1 - 2 X + 6 i T -(X-i T)^3/3 \cr
 -1    & (X+i T)^2 \cr
\end{pmatrix}.
\end{eqnarray}
After permuting the rows $N-1$ and $N$, 
the upper left block of order $N-1$ (in this simple example a scalar) is upper triangular and nonsingular,
making the formula (\ref{eq-det-blocks}) applicable.
The final result is a determinant of order $N-1$ in the $2N-2$ variables $X,T,\apib_{1:N-1},\amib_{1:N-3}$
(instead of a Wronskian of order $2 N$ in the $2 N$ variables $X,T,\apib_{1:N-1},\amib_{1:N-1}$),
in this simple example just a scalar,
\begin{eqnarray}
& & {\hskip -8.0truemm}
\det \partial_{\amib_{1}} M_1 = \Norm_1^2 \det\begin{pmatrix}\Rsj_1\end{pmatrix}
=\Norm_1^2 \Rsj_1.
% -(3*I)*T-(1/3*I)*T^3+X*T^2+I*T*X^2-(1/3)*X^3+X+s1
%4 \left[\apib - X^3/3 - i T^3/3 + X T^2 +i X^2 T - 3 i T\right].
\label{eqdetfinal}
\end{eqnarray}

In order to estimate the gain of the present algorithm,
the four polynomials (\ref{eq-def-squares}) 
for $N=6$ in 5 arbitrary complex parameters are obtained in 10 minutes in a laptop,
those for $N=7$ in 10 arbitrary complex parameters in 60 minutes in a desktop,
those for $N=10$ and zero values of all parameters in 20 minutes in a laptop.

Full details can be found
in Appendix \ref{Appendix-First-few}
and in the supplementary material, see section \ref{Supplementary-material}.

% ============================================================================
\section{On recurrence relations for NLS} 
\label{section-other-recurrences}

Let us denote in this section $v_j(X,T)$ various solutions of the NLS equation (\ref{eqNLS}),
not necessarily rational.

There exist at least two other kinds of recurrence relations among solutions of NLS,
and both are four-term ones. 

The first kind is the NLSF.
By definition \cite{Lamb1974,Chen1974}, the NLSF of a soliton equation 
is obtained by an elimination process between four copies of the B\"acklund transformation,
involving two spectral parameters $\lambda_1, \lambda_2$
and four solutions $v_0(X,T)$, $v_{1,1}=v_1(X,T,\lambda_1)$, $v_{1,2}=v_1(X,T,\lambda_2)$, $v_2(X,T,\lambda_1, \lambda_2)$.
In the case of NLS,
the result \cite{RogersShadwick}
is a set of two algebraic (not involving derivatives) equations 
in these four solutions and their complex conjugates,
\begin{eqnarray}
& & 
% Xiao Yi (4)
+(v_0-v_{1,1}) \sqrt{\lambda_1^2-|v_0+v_{1,1}|^2}
-(v_0-v_{1,2}) \sqrt{\lambda_2^2-|v_0+v_{1,2}|^2}
\nonumber\\ & & 
-(v_2-v_{1,2}) \sqrt{\lambda_1^2-|v_2+v_{1,2}|^2}
-(v_2-v_{1,1}) \sqrt{\lambda_2^2-|v_2+v_{1,1}|^2}=0,
\label{eq-NLS-BT-algebraic}
\end{eqnarray}
and it defines $v_2$ as the root of a very high degree polynomial equation, 
therefore not practicable at all.

\textit{Remark}.
The nice expression \cite[Eq.~(13)]{Xiao-1991-NLS} 
	of $v_2-v_0$ as a rational function of $v_{1,1}$, $v_{1,2}$
	and the four square roots of (\ref{eq-NLS-BT-algebraic})
	only applies to the sequence (vacuum, one-soliton, one-soliton, two-soliton).

The recurrence of the second kind is a reduction of the Toda chain,
and its four elements, now denoted $v_0, v_1, v_2, v_3$, are not contiguous elements of the B\"acklund transformation.
It expresses $v_3$ \cite[Eqs.~(1.8ab)]{CLZ1988} \cite[Eq.~(8)]{Adler-Yamilov-1994-NLS}
as a rational function of three other solutions
and their complex conjugates,
\begin{eqnarray}
& & 
% CLZ q3=q0+(k-h) q2/(1-q2 r1/4), r3=r0+(h-k) r1/(1-q2 r1/4) (very dissymmetric)
     v_3=     v_0 + \lambda \frac{     v_2}{1-v_2 \bar v_1},
\bar v_3=\bar v_0 - \lambda \frac{\bar v_1}{1-v_2 \bar v_1},
\label{eq-reduction-CLZ}
\end{eqnarray}
and it depends on only one spectral parameter instead of two.
If one assigns $v_0, v_1, v_2$ to the rational rogue waves 
$u_1$, $u_2(\alpha_1,\beta_1)$, $u_3(\alpha_1,\beta_1,\alpha_2,\beta_2)$,
then $v_3$ is the quotient of two polynomials of degrees $20$ and $20$, as expected,
but the condition that $v_3$ obeys NLS implies $\lambda=0$.

% ============================================================================
\section{Other representations of the rational $N$-th wave} 
\label{section-Other}

The difference between $u_N$ and a lower wave is also a bilinear rational function
of the two complex parameters $\alpha_{N-1}$ and $\beta_{N-1}$,
this time with different canonical affine factors
\begin{eqnarray}
& & (j \not=N-2):\ 
% u2RW:=u1RW+exp(I*T/2)*(Ca1^2+Cb1^2+Cc1^2)/(G2d*G1d)/(1+I*T);
%u_N=u_1+\frac{C(a_1)^2+C(b_1)^2+\Ccone^2}{(1+i T) \Den_1 \Den_2} e^{i T/2},
\frac{u_N}{u_0}=\frac{u_{j}}{u_0}+\frac{\Csj_{N-1,j} \Cdj_{N-1,j}+\nu_N \Nrj_{N-1}}{\Den_N\Den_{j}},
\label{eq-uN-minus-uj} 
\end{eqnarray}
and the choice $j=0$ (subtraction of the plane wave) commonly adopted by many authors
has two advantages:
its denominator is the same as that of $u_N$,
and its numerator has a lower degree in $X$ and $T$.
However, it also has an inconvenient:
the size of the canonical affine factors $\Csj_{N-1,j}$ is bigger than those of  $\Csj_{N-1}$.

But there is a remarkable fact:
when one subtracts the second previous wave $j=N-2$,
the new numerator is no more bilinear but linear separately in $\alpha_{N-1}$ and $\beta_{N-1}$,
\begin{eqnarray}
& & 
% u2RW:=u1RW+exp(I*T/2)*(Ca1^2+Cb1^2+Cc1^2)/(G2d*G1d)/(1+I*T);
%u_N=u_1+\frac{C(a_1)^2+C(b_1)^2+\Ccone^2}{(1+i T) \Den_1 \Den_2} e^{i T/2},
\frac{u_N}{u_0}=\frac{u_{N-2}}{u_0}+\frac{\lambda_N \Csj_{N-1,N-2}+\mu_N \Cdj_{N-1,N-2}+\nu_N \Nrj_{N-1}}{\Den_N\Den_{N-2} P_N},
\label{eq-uN-minus-uNminustwo} 
\end{eqnarray}
and then the sizes of $\Csj_{N-1,N-2}$ and $\Cdj_{N-1,N-2}$ are the smallest of the sets
$\Csj_{N-1,j}$ and $\Cdj_{N-1,j}$, $j\le N-2$.
The reason could be an underlying nonlinear superposition formula,
close to the formula (\ref{eq-reduction-CLZ}),
and whose expression still has to be uncovered.

% ============================================================================
\section{Conclusion}
\label{Conclusion}

The present very compact explicit expressions 
 open the possibility to investigate
the existence of new patterns in addition to the already observed ones (concentric
rings, polygonal configurations, . . . ).

The present procedure could help to solve the difficulty encountered in the vector
NLS system
\cite[page 41]{ZPFW-Vector-NLS}.

% ============================================================================
\section{Supplementary material}
\label{Supplementary-material}

These are totally five files :
one file containing a Maple program, plus
four data files in Maple input format.
The data files contain,
for $N$ from 0 to 7, real and imaginary parts of 
the polynomials (\ref{eq-def-squares}),
as well as:
their number of terms, degree in $X$ and $T$, list of contributing parameters
(\ref{eq-def-param}),
degree in each parameter,
sometimes the time and storage needed to compute them.
Due to their various sizes, they are split into several files.
\begin{enumerate}
    \item File ``1-4.data'', small,       $N$ from 0 to 4. 
    \item File   ``5.data'', medium,      $N=5$, eight  arbitrary real parameters.
    \item File   ``6.data'', large,       $N=6$, ten    arbitrary real parameters.
    \item File   ``7.data'', very large,  $N=7$, twelve arbitrary real parameters.
\end{enumerate}

The number of terms of polynomials 
$\Raj_j$, $\Rbj_j$, $\Caj_j$, $\Cbj_j$, $j=N-1$,
as defined and available in Supplementary material files *.data, is the following.

%$N=2$, number of lines is 1, 1, 1, 1.
%$N=3$, number of lines is 3, 3, 5, 6.
%$N=4$, number of lines is 32, 29, 49, 50.
%$N=5$, number of lines is 372, 375, 696, 710.
%$N=6$, number of lines 5119, 5165, 9730, 9823.
%$N=7$, number of lines 74987, 86029, 143146, 143329.

$N=2$, number of terms 4, 4, 5, 7.

$N=3$, number of terms 24, 14, 37, 42.

$N=4$, number of terms 189, 188, 319, 333.

$N=5$, number of lines 1846, 1845, 3271, 3.313

$N=6$, number of terms 19086, 19080, 34717, 34838.

$N=7$, number of terms 203493, 203532, 375625, 376043.

Appendix
\ref{Appendix-First-few} contains all the results up to $N=3$ included.

% ============================================================================
\section{Acknowledgments}

The author warmy thanks Pierre Gaillard for sharing his computer algebra program.
\vfill\eject

% ****************************************************** References

\section*{Data availability statement}

Data available in article or supplementary material.
The data that support the findings of this study are available within the article [and its supplementary material].

The author has no conflicts to disclose.
\vfill\eject 

% =============================================================================
\appendix

\section{The first few rational waves}
\label{Appendix-First-few}

Given the plane wave $u_0=e^{i T/2}$,
the $N$-th rational rogue wave of NLS, Eq.~(\ref{eqNLS}),
is
\begin{eqnarray}
& & 
\frac{u_N}{u_0}=\frac{\Num_N(X,T)}{\Den_N(X,T)}, \Num_N(X,T)=\det M_3, \Den_N(X,T)=\det M_1,
%\label{eq-N-0-1} 
%\nonumber \hfill (\ref{eq-N-0-1}) 
\end{eqnarray}
with the normalization $\Den_N(0,0)=1$ when all the $2 N-2$ real parameters $a_j$, $b_j$ vanish.

Our notation is $C$ like complex (numerator), $R$ like real (denominator).

The plane wave is
\begin{eqnarray}
& & u_0=e^{i T/2}, \Num_0=1, \Den_0=1
\end{eqnarray}
and the first (Peregrine) wave is
\begin{eqnarray}
%u1RWter:=(1+X**2+(T-2*I)**2)/(1+X^2+T^2) *exp(I*T/2);
& & {\hskip-10.0truemm}
\Num_1=1+X^2+(T-2 i)^2, \Den_1=1+X^2+T^2, \Num_1(0,0)=-3, \Den_1(0,0)=1.
\end{eqnarray}

The shortest expressions which define the second wave $N=2$ are
\begin{eqnarray}
& & {\hskip -10.0truemm}
\left\lbrace
\begin{array}{ll}
%Ra1:=a1-X^3/3+X*(1+T**2):
%Rb1:=b1-T^3/3-T*(3-X**2):
%Ca1:=a1-X^3/3+X*(1+(T-2*I)^2):
%Cb1:=b1-T^3/3+(T-2*I)*X^2+(T-2*I+2*I*T^2):
    \displaystyle{\Raj_1=a_1-\frac{X^3}{3}+ X (1+T^2), 
}\\ \displaystyle{\Rbj_1=b_1-\frac{T^3}{3}- T (3-X^2),
}\\ \displaystyle{\Caj_1=a_1-\frac{X^3}{3}+ X (1+(T-2 i)^2), 
}\\ \displaystyle{\Cbj_1=b_1-\frac{T^3}{3}+(T-2 i) X^2 + (T-2 i + 2 i T^2),
}\\ \displaystyle{\Rsj_1=\Raj_1+i \Rbj_1, \Rdj_1=\overline{\Rsj_1},
\Csj_1=\Caj_1+i \Cbj_1,
\Cdj_1=\Caj_1-i \Cbj_1.
%Rs1:=subs(rtosd,Ra1+I*Rb1):
%Rd1:=subs(rtosd,Ra1-I*Rb1):
%Cs1:=subs(rtosd,Ca1+I*Cb1):
%Cd1:=subs(rtosd,Ca1-I*Cb1):
}
\end{array}
\right.
\label{eq-u2RW} 
\end{eqnarray}

The third wave $N=3$ is characterized by
\begin{comment}

Ra2:=Den1*a2+X*(a1^2-b1^2)-2*T*a1*b1
 +( 2/3*X*T^3+2/3*T*X^3+6*X*T)*b1+(-1/3*T^4+1/3*X^4+3/2*X^2+7/2*T^2-3/2)*a1
 +X*(-X^6/45+(T^2+1)*X^4/5+(1-T^2/3)^2*X^2-T^6/9-3*T^4-11*T^2-1):

Rb2:=Den1*b2+T*(a1^2-b1^2)+2*X*a1*b1
 +(-2/3*X*T^3-2/3*T*X^3-6*X*T)*a1+( 1/3*X^4-1/3*T^4+3/2*X^2+7/2*T^2-3/2)*b1
 +T*( T^6/45+(-X^2/5+17/15)*T^4+(-1-2*X^2-X^4/9)*T^2+X^6/9-X^4-5*X^2+5):

Ca2:=Num1*a2+X*(a1^2-b1^2)-2*(T-2*I)*a1*b1
 +(1/3*X^4+8/3*I*T^3-1/3*T^4+1/2-14*I*T+23/2*T^2+3/2*X^2)*a1
 +(2/3*T*X^3+2/3*X*T^3-2*X*T+4*I*X-4/3*I*X^3-4*I*T^2*X)*b1
 +X*(-X^6/45+(T^2-4*I*T-3)*X^4/5+(T^4-8*I*T^3-30*T^2+24*I*T-15)*X^2/9
     -T^6/9+4*I*T^5/3+11*T^4/3+8*I*T^3/3+13*T^2-20*I*T-5):

Cb2:=Num1*b2+(T-2*I)*(a1^2-b1^2)+2*X*a1*b1
 +(1/3*X^4+8/3*I*T^3-1/3*T^4+1/2-14*I*T+23/2*T^2+3/2*X^2)*b1
 +(-2/3*T*X^3+4/3*I*X^3-4*I*X+4*I*T^2*X-2/3*X*T^3+2*X*T)*a1
 +T^7/45-14*I*T^6/45-(11+3*X^2)*T^5/15+I*(-10/3+2*X^2)*T^4
 +(-X^4/9+6*X^2-59/3)*T^3+I*((2/3)*X^4-4*X^2+22)*T^2
 +(X^6/9+X^4/3+3*X^2+1)*T+I*(-(2/9)*X^6-(2/3)*X^4-6*X^2-2):
\end{comment}  

\begin{eqnarray}
& & {\hskip -10.0truemm}
\left\lbrace
\begin{array}{ll}
    \displaystyle{\Raj_2=\Den_1 a_2 +(a_1^2-b_1^2) X - 2 a_1 b_1 T
}\\ \displaystyle{  \phantom{12345}   
  +a_1(X^4/3-T^4/3+3/2 X^2+7/2 T^2-3/2 )    
}\\ \displaystyle{   \phantom{12345} 
  +b_1 (2/3 X T^3+2/3 T X^3+6 X T)
}\\ \displaystyle{  \phantom{12345}   
   +X (-X^6/45+(T^2+1) X^4/5+(1-T^2/3)^2 X^2-T^6/9-3 T^4-11 T^2-1), %
}\\ \displaystyle{\Rbj_2=\Den_1 b_2+(a_1^2-b_1^2) T+2 a_1 b_1 X
}\\ \displaystyle{  \phantom{12345}    
 +b_1 ( X^4/3-T^4/3+3/2 X^2+7/2 T^2-3/2)
}\\ \displaystyle{  \phantom{12345}   
 +a_1 (-2/3 X T^3-2/3 T X^3-6 X T)
}\\ \displaystyle{  \phantom{12345}    
 +T \left[T^6/45+(-X^2/5+17/15) T^4+(-1-2 X^2-X^4/9) T^2+X^6/9-X^4-5 X^2+5\right],
}\\ \displaystyle{\Caj_2=\Num_1 a_2+(a_1^2-b_1^2)X-2 a_1 b_1 (T-2 i)
}\\ \displaystyle{  \phantom{12345}    
 +a_1 (X^4/3+8/3 i T^3-T^4/3+1/2-14 i T+23/2 T^2+3/2 X^2)
 }\\ \displaystyle{  \phantom{12345}     
 +b_1 (2/3 T X^3+2/3 X T^3-4/3 i X^3-4 i T^2 X-2 X T+4 i X)
 }\\ \displaystyle{  \phantom{12345}    
 +X \left[-X^6/45+(T^2-4 i T-3) X^4/5+(T^4-8 i T^3-30 T^2+24 i T-15) X^2/9
\right.}\\ \displaystyle{\left.  \phantom{12345}    
     -T^6/9+4 i T^5/3+11 T^4/3+8 i T^3/3+13 T^2-20 i T-5\right],
}\\ \displaystyle{\Cbj_2=\Num_1 b_2+(a_1^2-b_1^2)(T-2 i)+2 a_1 b_1 X
}\\ \displaystyle{  \phantom{12345}    
 +b_1 (X^4/3+8/3 i T^3-T^4/3+1/2-14 i T+23/2 T^2+3/2 X^2)
 }\\ \displaystyle{  \phantom{12345}     
 +a_1 (-2/3 T X^3-2/3 X T^3+4/3 i X^3+4 i T^2 X+2 X T-4 i X)
 }\\ \displaystyle{  \phantom{12345}    
 +T^7/45-14 i T^6/45-(11+3 X^2) T^5/15+i (-10/3+2 X^2) T^4
 }\\ \displaystyle{  \phantom{12345}    
 +(-X^4/9+6 X^2-59/3) T^3+i ((2/3) X^4-4 X^2+22) T^2
}\\ \displaystyle{ \phantom{12345}     
 +(X^6/9+X^4/3+3 X^2+1) T+i (-(2/9) X^6-(2/3) X^4-6 X^2-2),
}\\ \displaystyle{\Rsj_2=\Raj_2+i \Rbj_2, \Rdj_2=\overline{\Rsj_2},
\Csj_2=\Caj_2 + i \Cbj_2,
\Cdj_2=\Caj_2 - i \Cbj_2.
%Rs2:=subs(rtosd,Ra2+I*Rb2):
%Rd2:=subs(rtosd,Ra2-I*Rb2):
%Cs2:=subs(rtosd,Ca2+I*Cb2):
%Cd2:=subs(rtosd,Ca2-I*Cb2):
}
\end{array}
\right.
\label{eq-u3RW} 
\end{eqnarray}

Fo the elements of the next waves up to $N=7$ included,
see the section \ref{Supplementary-material}.

\vfill\eject
%\end{CJK}
\end{document}